\documentstyle[12pt,epsf]{article}
\def\photonatomrightt{\begin{picture}(3,1.5)(0,0)
                               \put(0,-0.75){\tencirc \symbol{2}}
                               \put(1.5,-0.75){\tencirc \symbol{1}}
                               \put(1.5,0.75){\tencirc \symbol{3}}
                               \put(3,0.75){\tencirc \symbol{0}}
                     \end{picture}
                    }

\def\photonatomright{\begin{picture}(3,1.5)(0,0)
                                \put(0,-0.75){\tencircw \symbol{2}}
                                \put(1.5,-0.75){\tencircw \symbol{1}}
                                \put(1.5,0.75){\tencircw \symbol{3}}
                                \put(3,0.75){\tencircw \symbol{0}}
                      \end{picture}
                     }

\def\photonatomup{\begin{picture}(1.5,3)(0,0)
                             \put(-0.75,3){\tencircw \symbol{3}}
                             \put(-0.75,1.5){\tencircw \symbol{2}}
                             \put(0.75,1.5){\tencircw \symbol{0}}
                             \put(0.75,0){\tencircw \symbol{1}}
                   \end{picture}
                  }

\def\photonrightthalf{\begin{picture}(15,1.5)(0,0)
                    \multiput(0,0)(3,0){5}{\photonatomrightt}
                 \end{picture}
                }

\def\photonrighthalf{\begin{picture}(30,1.5)(0,0)
                     \multiput(0,0)(3,0){5}{\photonatomright}
                  \end{picture}
                 }

\def\photonuphalf{\begin{picture}(1.5,15)(0,0)
                      \multiput(0,0)(0,3){5}{\photonatomup}
                   \end{picture}
                  }

\def\fermionurr{\begin{picture}(30,15)(0,0)
                        \put(-30,-15){\vector(2,1){15}}
                        \put(-15,-7.5){\line(2,1){15}}
                  \end{picture}
                 }
\def\fermionurrhalf{\begin{picture}(15,7.5)(0,0)
                        \put(-15,-7.5){\vector(2,1){7.5}}
                        \put(-7.5,-3.75){\line(2,1){7.5}}
                  \end{picture}
                 }
\def\fermiondrr{\begin{picture}(30,15)(0,0)
                        \put(0,0){\vector(2,-1){15}}
                        \put(15,-7.5){\line(2,-1){15}}
                  \end{picture}
                 }
\def\fermiondrrhalf{\begin{picture}(15,7.5)(0,0)
                        \put(0,0){\vector(2,-1){7.5}}
                        \put(7.5,-3.75){\line(2,-1){7.5}}
                  \end{picture}
                 }

\newenvironment{Feynman}[3]{\begin{center}
                            \setlength{\unitlength}{#3 mm}
                            \begin{picture}(#1)(#2)
                            \thicklines
                           }{\end{picture} \end{center}}

\renewcommand{\cal}{\mathcal}
\newcommand{\litwo}{ \mbox{Li}_2 }
\newcommand {\oalf} {\mbox{${\cal O}(\alpha)$}}
\newcommand{\nn}{\noindent}
\newcommand{\nl}{\nonumber \\}
\newcommand{\bq}{\begin{equation}}
\newcommand{\eq}{\end{equation}}
\newcommand{\ba}{\begin{eqnarray}}
\newcommand{\ea}{\end{eqnarray}}

\newcommand {\ql} {\mbox{$Q^2_{l}  $}}
\newcommand {\yl} {\mbox{$y  _{l}  $}}
\newcommand {\xl} {\mbox{$x   _{l}  $}}

\newcommand {\yh} {\mbox{$y  _{h}  $}}

\newcommand {\qh} {\mbox{$Q^2_{h}  $}}

\newcommand {\xjb} {\mbox{$x_  {_{\mathrm{JB}}}$}}

\newcommand {\qjb} {\mbox{$Q^2_{_{\mathrm{JB}}}$}}

\newcommand {\ysi} {\mbox{$y_{_{\mathrm{\Sigma}}}$}}
\newcommand {\qsi} {\mbox{$Q^2_{_{\mathrm{\Sigma}}}$}}

\newcommand {\yesi} {\mbox{$y_{_{\mathrm{e\Sigma}}}$}}

\newcommand {\ydo} {\mbox{$y_{_{\mathrm{DA}}}$}}
\newcommand {\qdo} {\mbox{$Q^2_{_{\mathrm{DA}}}$}}

\newcommand {\xan} {\mbox{$x_  {\theta \mathrm{y}}$}}

\newcommand {\qan} {\mbox{$Q^2_{\theta \mathrm{y}}$}}

\begin{document}
\thispagestyle{empty}
\onecolumn
\vspace{-1.4cm}
\begin{flushleft}
{DESY 95--085 \\}
{hep-ph/9504423\\}
revised
\\
July 1995
\end{flushleft}
\vspace{1.5cm}
\begin{center}
{\LARGE  {
Complete $\cal O$($\alpha$) QED corrections
\vspace*{.2cm} \\
to the process $ep \rightarrow eX$
in mixed variables
\vspace*{1.0cm}
}}
\end{center}
\nn
{\large
Dima Bardin$^{1,2 }$, \hfill 
Pena Christova$^{3\#}$, \hfill 
Lida Kalinovskaya$^{2\#}$ \hfill
and $\;$ Tord~Riemann$^1$}
\\
\vspace*{1.0cm}

\begin{itemize}
\item[1]
DESY -- Zeuthen, Platanenallee 6, D--15738 Zeuthen, Germany
\item[2]
Bogoliubov Laboratory for Theoretical Physics, JINR,
ul. Joliot-Curie 6, 
RU--141980 Dubna,
Moscow Region, Russia
\item[3]
Dept. of Physics, Faculty of Physics,
Higher Pedagogical Institute, Shoumen, Bulgaria
\end{itemize}

\vfill
\vspace*{1.0cm}

\nn
{\large Abstract}
\vspace*{.3cm}
\\
\nn
The complete set of ${\cal O}(\alpha)$ QED corrections with soft photon
exponentiation to the process $ep \rightarrow eX$
in mixed variables ($y=y_h,Q^2=Q_l^2$) is calculated in the quark
parton model, including the lepton-quark interference and the
quarkonic corrections which were unknown so far.
The interference corrections amount to few
percent or less and become negligible at small $x$.
The leading logarithmic terms proportional to
$\ln(Q^2/m_q^2)$ from radiation off quarks
are discussed and the non-logarithmic quarkonic corrections found to be
negligible for almost all experimentally accessible $x$ and $y$.
\vspace*{1.0cm}
\vfill
\footnoterule
\nn
{
{
{\footnotesize
Emails:
bardindy@cernvm.cern.ch, penka@uni-shoumen.bg,
lika@thsun1.jinr.dubna.su, riemann@ifh.de
\\
$^{\#}$
Supported by EU contract CHRX--CT--92--0004.
}
}
}
\pagebreak

\section{
\label{sin}
Introduction}
Traditionally, deep inelastic $ep$ scattering,
\ba
e(k_1) + p(p_1) \rightarrow e(k_2) + X(p_2),
\label{e1}
\ea
accompanied by radiative corrections from the reaction
\ba
e(k_1) + p(p_1) \rightarrow e(k_2) + X(p_2) + n\gamma(k),
\label{e2}
\ea
was studied in terms of kinematic variables, which could be determined
exclusively from the electron.
The new detector generation at HERA gives additional access to the hadron
kinematics.
For the radiative process~(\ref{e2})
the `leptonic' variables differ from those, which are determined
partly or exclusively with information about the hadronic kinematics.
QED corrections from the diagrams of figure~\ref{fig5}
(henceforth called leptonic corrections)  with soft
photon exponentiation
have been determined in a model independent approach in several
variables~\cite{MI,Hector}.
The same has been undertaken in leading logarithmic approximation,
including higher order corrections~\cite{Hector,jb}, for a larger
variety of variable sets~\cite{variables}.
For a complete ${\cal O}(\alpha)$ calculation, one has to
use the quark-parton model and include
corrections from the diagrams of figure~\ref{fig6}
and from their interferences with those of figure~\ref{fig5}.
In terms of leptonic variables, this has been done some time
ago~\cite{disepl}.

In this letter, we report on the corresponding corrections in terms of
{\em mixed} variables.
These variables are defined as follows:
\ba
y_m   \equiv y_h   = \frac{-2p_1Q_h}{S},
\hspace{1.5cm} 
Q_m^2 \equiv Q_l^2 = (k_1-k_2)^2,
\hspace{1.5cm} 
x_m   \equiv \frac{Q_m^2}{y_m S}.
\label{e3}
\ea

In the next section, we explain the structure of the corrections and
give explicit expressions for them.
Section 3 contains a discussion of photonic bremsstrahlung from quarks
and section~4 numerical results.
The technically involved calculations will be described elsewhere.

\vspace{2mm}

\section{The bremsstrahlung corrections}
The cross section of reaction (2) may be written as follows\footnote
{If not stated differently, the variables $x,y,Q^2$ are {\em mixed}
  variables.}:
\ba
\frac{d^2 \sigma}{dxdy}
&=&
\frac{2\pi \alpha^2 S x}{Q^4}
\, \ \sum_{Q,\bar Q}
\Biggl\{
B_0(y,1)  f_Q(x)
\left[ 1+ \frac{\alpha}{\pi} \sum_{a=e,i,q} c_aS_a(x,y|m_a^2) \right]
\nl &&+~
\frac{\alpha}{\pi} \,\,  \sum_{a=e,i,q}
\Biggl[
\int_1^{1/x}dz \left[{\bar B}_a^V(z;x,y)+  p_ep_Q {\bar B}_a^A(z;x,y)\right]
\nl &&
\hspace*{2.0cm}
+~
\int_y^1dz \left[B_a^V(z;x,y)+  p_ep_Q B_a^A(z;x,y)\right]
 \Biggr]
\Biggr\}
+
\frac{d^2 \sigma^{box}}{dxdy},
\label{e4}
\ea
with
\ba
B_0(y,z) &=&  V_0 \, Y_{+}\left(\frac{y}{z}\right) + p_ep_Q A_0 \,
Y_{-}\left(\frac{y}{z}\right)
\label{byz}
\ea
and $Y_{\pm}(y)=1 \pm (1-y)^2$.

The sum over $Q$ and $\bar Q$ extends over the quark content of the
proton.
The second sum with $a=e,q,i$ extends over the photonic corrections
originating from the lepton legs, from the quark legs, and from their
interference.
Correspondingly, $c_e=Q_e^2, c_q=Q_Q^2, c_i=Q_eQ_Q$. We use the
convention $Q_e=-1$.
The couplings $Q_f, v_f, a_f$ are collected in overall factors of the
vector type,
$V(B,p)$, and axial vector type,
$A(B,p)$
and the running coupling
$\alpha_{_{QED}}(Q^2)$,
the
Fermi constant $G_F$, and the ratio of
the photon and $Z$ propagators in effective coupling strengths
$K(B,p)$:

\begin{figure}[th]
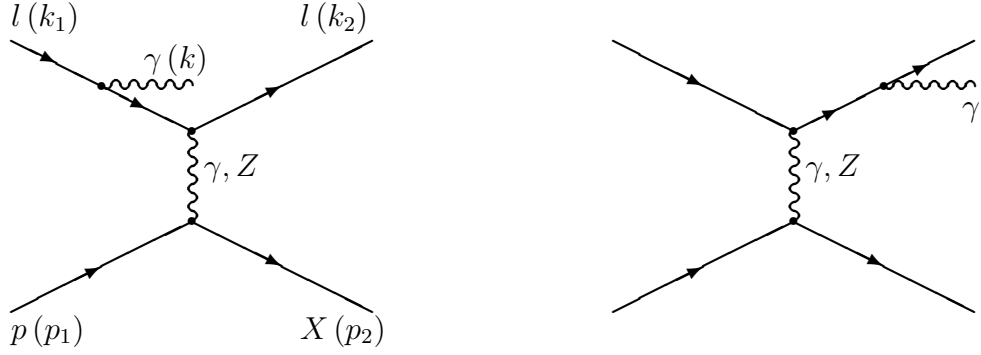

\vspace{1.5cm}
\begin{minipage}[th]{16cm}{
\begin{center}
\begin{Feynman}{160,65}{0,0}{0.8}
\put(00,074.5){\fermiondrrhalf}
\put(00,077){$l \, (k_1)$}
\put(48,077){$l \, (k_2)$}
\put(32,052){$\gamma,Z$}
\put(22,070){$\gamma \, (k)$}
\put(60,074.5){\fermionurr}
\put(15,067){\fermiondrrhalf}
\put(30,059,5){\circle*{1,5}}
\put(30,044,5){\circle*{1,5}}
\put(30,044,5){\fermiondrr}
\put(30,044,5){\fermionurr}
\put(48,25){$X \, (p_2)$}
\put(00,25){$p \, (p_1)$}
\put(15,067){\photonrighthalf}
\put(30,044,5){\photonuphalf}
\put(15,067){\circle*{1,5}}
\put(100,074.5){\fermiondrr}
\put(160,074.5){\fermionurrhalf}
\put(132,052){$\gamma, Z$}
\put(158,062){$\gamma$}
\put(145,067){\circle*{1,5}}
\put(145,067){\photonrighthalf}
\put(145,067){\fermionurrhalf}
\put(130,059,5){\circle*{1,5}}
\put(130,044,5){\circle*{1,5}}
\put(130,044,5){\photonuphalf}
\put(130,044,5){\fermiondrr}
\put(130,044,5){\fermionurr}
\end{Feynman}
\end{center}
}\end{minipage}
\vspace*{-1.5cm}
\caption{\it
Leptonic bremsstrahlung.
\label{fig5}
}
\end{figure}

\begin{figure}[t]
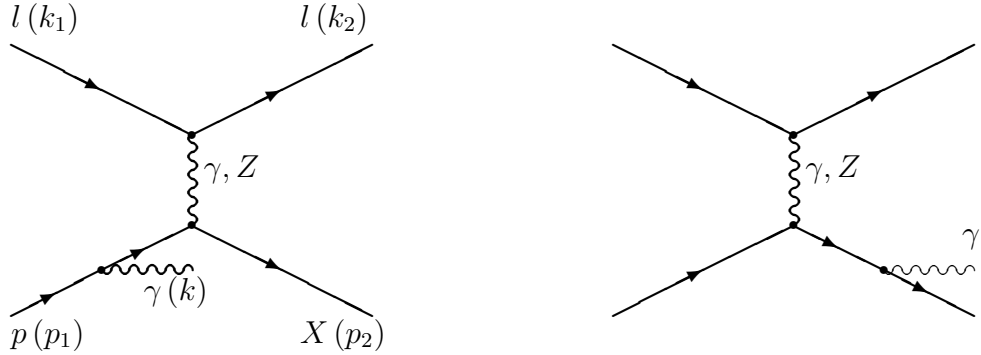

\vspace{0.9cm}
\begin{minipage}[t]{16cm}{
\begin{center}
\begin{Feynman}{160,65}{0,0}{0.8}
\put(00,55){\fermiondrr}
\put(00,58){$l \, (k_1)$}
\put(48,58){$l \, (k_2)$}
\put(32,33){$\gamma, Z $}
\put(22,12,5){$\gamma \, (k)$}
\put(60,55){\fermionurr}
\put(30,40){\circle*{1,5}}
\put(30,25){\circle*{1,5}}
\put(30,25){\fermiondrr}
\put(30,25){\fermionurrhalf}
\put(15,17.5){\fermionurrhalf}
\put(48,5.5){$X \, (p_2)$}
\put(00,5.5){$p \, (p_1)$}
\put(30,25){\photonuphalf}
\put(15,17.5){\photonrighthalf}
\put(15,17.5){\circle*{1,5}}
\put(100,55){\fermiondrr}
\put(132,33){$\gamma, Z$}
\put(160,55){\fermionurr}
\put(130,40){\circle*{1,5}}
\put(130,25){\circle*{1,5}}
\put(130,25){\photonuphalf}
\put(145,17.5){\photonrightthalf}
\put(158,22){$\gamma$}
\put(145,17.5){\circle*{1,5}}
\put(145,17.5){\fermiondrrhalf}
\put(130,25){\fermiondrrhalf}
\put(130,25){\fermionurr}
\end{Feynman}
\end{center}
}\end{minipage}
\caption{\it
Bremsstrahlung from quarks.
\label{fig6}
}
\vspace*{1.5cm}
\end{figure}

\ba
V_0&=&V(1,1),
\nl
V(a,b)&=& \sum_{B=\gamma,I,Z} K(a,b;B,p_ep_Q)\,  V(B,p_ep_Q),
\\
K(a,b;B,p)&=&\chi_{B_1}(a) \, \chi_{B_2}(b) \, K(B,p),
\nl
\chi_B(a)&=& \frac{Q^2}{aQ^2+M_B^2}.
\nonumber
\ea
Since the interaction of leptons with quarks proceeds via photon or
$Z$ boson exchange, we have three types of cross section pieces labeled
with $B=\gamma, Z, I$, where the latter denotes again the
interference.
The $[B_1 B_2]=\gamma\gamma, Z \gamma, ZZ$ correspond to
$B=\gamma,I,Z$.
The $V(B,p)$ and $K(B,p)$ contain weak one loop corrections; they are
defined in~\cite{disepl}\footnote{
Deviating from~\cite{disepl}, we took the $\chi_Z(1)$ out of the
definition of the $K(B,p)$.
An update of the weak corrections is described in~\cite{YR95}.
}.
The axial couplings $A(a,b)$ are related to the couplings $A(B,p)$
of~\cite{disepl} in the same manner as is explained above for the vector
couplings.
The sign factor is $p=p_ep_Q=+1$ for $PP$, $AA$ scattering and
$p=-1$ for $PA$ scattering, where we denote particles with $P$ and
anti-particles with $A$.

The functions $S_e,S_q$ contain factorizing soft photon corrections and
the corresponding vertex corrections\footnote
{
The exponentiation of the soft photon term $(\ln Q^2/m_e^2-1)\ln((1-y)(1-x)/x)$
is performed only for the leptonic
correction $S_e$~\cite{MI}.
}:  
\ba
S_e(x,y|m_e^2)
&=&
\left(\ln\frac{Q^2}{m_e^2} - 1 \right)
\ln\left[(1-y)\frac{1-x}{x}\right]
+\frac{3}{2}\ln\frac{Q^2}{m_e^2} - 2 - \ln(1-y)\ln(xy)
\nl
&&+~2 \, \litwo
(1) - \litwo (xy) + \litwo \left[-\frac{(1-x)y}{1-y} \right],
\\
S_q(x,y|m_Q^2)
&=&
S_e(x,y|m_Q^2)
-\left[
\litwo (1-y) + \litwo \left( - \frac{1-x}{x}\right)
\right] .
\label{e10}
\ea
With the exception of the last two terms in~(\ref{e10}) the expressions for
$S_e(x,y|m_e^2)$ and $S_q(x,y|m_Q^2)$ agree\footnote{
Here we treat the quark masses as independent
input parameters.
Instead, one could use the relation $m_q \equiv x_hM_p$,
where $M_p$ is the proton mass as it is suggested by kinematics.
Then, the square bracket in~(\ref{e10}) would change the sign.
In integrated expressions like~(\ref{e10}), it is then $m_q=x_mM_p$.
}.

The soft photon interference term
$S_i$ has to be combined with the $\gamma \gamma$ and $\gamma Z$ boxes in
order to get an infrared finite answer; the latter are added as an
extra piece after the divergences are canceled, and it is:
\ba
S_i(x,y)
&=&
\ln(1-y)\ln\frac{(1-y)^3(1-x)^2}{y^4} - 6 \, \litwo (y) + 2\, \litwo (xy) -
2 \, \litwo \left[-\frac{(1-x)y}{1-y} \right] .
\label{e9}
\ea

The corrections $B^{V,A}$ and ${\bar B}^{V,A}$ in~(\ref{e4}) are due to real
photonic bremsstrahlung.
Although we leave out all details of the involved calculations, a
remark on the phase space parameterization should be made.
The double-differential bremsstrahlung corrections are the result of a
threefold integration,
which has been organized such that two of the integrals may be
performed  analytically.
At the parton level, there are two nontrivial integrals to be
performed whose result has to be convoluted with the parton momentum
distribution:
$d \sigma^{brem} = \alpha^3/(2\pi S) \cdot
\sum_{Q,\bar Q}  f_Q(x_h) (dx_h/x_h)$ $dQ_l^2$\-$dQ_h^2$\-$ dy_l
d\Gamma_k |{\cal M}_{eq}^{brem}|^2$.
We chose the following parameterization of the phase space:
\ba
\frac{d^2\sigma^{brem}}{dxdy}
&=&
\frac{\alpha^3}{2\pi} Q_l^2
\sum_{Q,\bar Q}\, \int dz f_Q(z x) \int dy_l \, d\Gamma_k |{\cal
  M}_{eq}^{brem}|^2.
\label{e8}
\ea
Here, it is
\bq
z=\frac{x_h}{x_m} = \frac{Q_h^2}{Q_l^2},
\label{e31}
\eq
with $Q_h^2=(p_2-p_1)^2$.
The integration variable $z$ may be larger or smaller than 1.
As may be seen in figure~39 of~\cite{MI},
for these two cases the physical regions for the integration over
$y_l$ (at fixed values of
$Q_h^2$ or, equivalently, of $z$) are different, which in turn leads to
different integrands in~(\ref{e4}) for the subsequent integration over
$z$.
The integral over $y_l$ and that over the photon phase space,
indicated by $d\Gamma_k$, are performed analytically.
In fact the latter one is transformed into an integral over
$dz_2/(2\sqrt{R_z})$, where $z_2=-2kk_2$ is the denominator of the
final state electron propagator before an emission of a photon~\cite{MI}.

The hard bremsstrahlung functions in~(\ref{e4}) are:
\ba
{\stackrel {(-)}{B}}
_a^{V,A}(z;x,y)
&=&
\left\{ \begin{array}{c}V_a\\A_a\end{array}\right\}
\left\{
\sum_n Reg\left[ {\stackrel{(-)}{F}}_{an}^{V,A}(z,y) f_Q(zx)\right]
{\stackrel {(-)}{L}}_{an}  +
{\stackrel {(-)}{U}}_a^{V,A}(z,y)f_Q(zx)
\right\} .
\label{e11}
\ea
The bremsstrahlung vector couplings are:
\ba
V_e&=&c_e \, V(z,z), \hspace{0.7cm} V_i=c_i \, V(1,z), \hspace{0.7cm}
V_q=c_q \, V(1,1).
\label{e13}
\ea
The bremsstrahlung axial couplings are defined analogously.
The different arguments of $V(a,b)$ reflect the different momentum
flows in the exchanged boson propagators in figure~\ref{fig5}
($Q_h^2$) and figure~\ref{fig6} ($Q_l^2$).

For $a=e,i,q$, the $n$ runs from 1 to $2,1,2$.
The functions $F_{an}$ are multiplied by a factor $1/(1-z)$ which
becomes singular at $z=1$ due to the infrared singularity of the
corrections in this phase space region.
This singularity is regularized by a subtraction:
\ba
Reg\left[ {\stackrel{(-)}{F}}_{an}^{V,A}(z,y) f_Q(zx)\right]
&=&
\frac{y}{(1-z)}
\left[
{ \stackrel{(-)}{F}}_{an}^{V,A}(z,y) f_Q(zx)
-
{\stackrel{(-)}{F}}_{an}^{V,A}(1,y) f_Q(x)
\right]
{}.
\label{e12}
\ea
The quarkonic bremsstrahlung functions are:
\ba
F_{q1}^{V,A}(z,y) &=&
\frac{1}{2} \left(1+z^2\right)\frac{z}{y}
Y_{\pm}\left(\frac{y}{z}\right)
,
\\
F_{q2}^{V,A}(z,y) &=&
\frac{1}{2z^2} \left(1+z^2\right)\frac{1}{y}
Y_{\pm}\left( y \right)
,
\\
U_q^V &=&
\frac{1}{2} \left(2+3z+2z^2\right)
Y_{+}\left(\frac{y}{z}\right)
+\frac{1}{z}\left(1-\frac{y}{z}\right)
-\frac{y^2}{2z^2}\left(1+z^2\right)           \nl
&&-y(1+z)
- \frac{1}{2}+\frac{y^2}{z}\ln\frac{z}{y}
,
\\
U_q^A &=&
\frac{1}{2} \left(1+3z+2z^2\right)
Y_{-}\left(\frac{y}{z}\right)
-\frac{y^2}{2z^2}\left(1+z^2\right)-y(1+z)
,
\\
L_{q1}&=&L_1\left(z,y\left|\frac{m_Q^2}{z}\right.\right),
\label{lq1}
\\
L_1(z,y|m^2)&=& \ln \frac{Q^2}{m^2} - 1 + \ln\frac{z-y}{z(1-z)},
\\
L_{q2}(z,y)&=&
\ln \frac{1-z}{1-y}.
\label{e14}
\ea
The leptonic corrections differ by a simple factor:
\ba
\begin{array}{rclcrcl}
{\stackrel{(-)}{F}}_{ea}^{V,A} &=& z \,
{\stackrel{(-)}{F}}_{qa}^{V,A}, &\hspace{.0cm} a=1,2,
\hspace{.7cm}& {\stackrel{(-)}{U}}_e&=&z{\stackrel{(-)}{U}}_q,
\\
L_{e1}&=&L_1(z,y|m_e^2), &&  L_{e2}&=&L_{q2}.
\end{array}
\label{e15}
\ea
The lepton quark interference corrections have no mass singularities:
\ba
{\stackrel{(-)}{F}}_i^{V,A}&=& \frac{z}{y}
\left[
Y_{\pm}\left(\frac{y}{z}\right)+Y_{\pm}\left( y \right)
\right],
\hspace{1.cm}
 {\stackrel{(-)}{U}}_i^{V,A}=y{\stackrel{(-)}{\cal U}}_i^{V,A},
\\
L_i&=&\ln(1-y),
\\
{\cal U}_i^V&=&
2\left[
(2-y)\ln(1-y)-\left(4-y-\frac{y}{z}\right)\ln(1-z)-y\ln\frac{z}{y}\right]
-4(z-y),
\\
{\cal U}_i^A&=&
2 \, \frac{y}{z}\left[ \ln(1-y)-(1+z)\ln(1-z)-z\ln\frac{z}{y}\right] .
\label{e16}
\ea
The integrands in the two integration regions in~(\ref{e4})
are related by a symmetry relation:
\ba
{{\bar X}} \left( z,y,\frac{y}{z}, \ldots \right) &=& X \left(
\frac{1}{z},\frac{y}{z},y, \ldots \right),\hspace{0.7cm}
X=L_e,L_i,{\cal U}_i,F_q,L_q,U_q
{}.
\label{e17}
\ea

Finally, we have to discuss the last term in~(\ref{e4}), $d^2
\sigma^{box}/dxdy$.
This is the finite part of the contribution from the interference of
the $\gamma\gamma$
and $\gamma Z$ box diagrams with the Born diagrams.
It contributes to the factorizing part of the lepton quark interference.
Its variables are defined from the Born kinematics.
For this reason, we may refer to~\cite{disepl}, eqs.~(2.1), (3.1),
(3.4) and (C.11), (C.14).
The latter two formulae contain the corrections from the
box diagrams:
$S_i(B;a,b)^{box}=S_i(B;a,b)-S^{soft}$, $B=\gamma, Z$, where it is
$S^{soft}=4a^2\ln(b/a) \ln[(1-x)/x]$.

\section{Photonic bremsstrahlung from quarks}
\label{qisr}
In this section, we discuss the connection of our complete \oalf\
results with expressions to be expected from a LLA treatment.
For this purpose, we rewrite the terms proportional to $\ln(Q^2/m_Q^2)$
in the first of the integrals in~(\ref{e4}) and perform a change of
integration variable, $z \rightarrow 1/z$:
\ba
\frac{d^2 \sigma_{LLA,q}^{ini}}{dxdy}
&=&
\frac{2\pi \alpha^2 S x}{Q^4}  \sum_{Q,\bar Q} \;
\frac{\alpha}{\pi} Q_Q^2 \ln\frac{Q^2}{m_Q^2} B_0(y,1)
\int_x^1dz \left[ \frac{1+z^2}{1-z} \; \frac{1}{2z}\;
 f_Q\left(\frac{x}{z}\right) -  \frac{1}{z(1-z)} \;  f_Q(x) \right] .
\nl
\label{e4a}
\ea
In the second integral, one has to collect terms correspondingly (here
without change of integration variable):
\ba
\label{e4b}
\frac{d^2 \sigma_{LLA,q}^{fin}}{dxdy}
&=&
\frac{2\pi \alpha^2 S x}{Q^4}   \sum_{Q,\bar Q}  \;
\frac{\alpha}{\pi} Q_Q^2 \ln\frac{Q^2}{m_Q^2} \int_y^1dz
\left[ \frac{1+z^2}{1-z} \; \frac{z}{2}   \;  f_Q(z x)B_0(y,z)
-  \frac{1}{1-z} \;  f_Q(x)B_0(y,1) \right].
\nl
\ea
Further, the soft photon correction~(\ref{e10}) has to be rewritten:
\ba
S_q(x,y|m_Q^2)
\equiv
S_q^{ini}(x|m_Q^2)
+
S_q^{fin}(y|m_Q^2)
+
S_q^{finite}(x,y),
\label{smod}
\ea
where
\ba
S_q^{ini}(x|m_Q^2)
&=&
\ln\frac{Q^2}{m_Q^2}
\left[
 -\frac{1}{2}\int_0^1dz \frac{1+z^2}{1-z}
+ \int_x^1dz \frac{1}{z(1-z)}
\right]
,\label{smodini}
\\ 
S_q^{fin}(y|m_Q^2)
&=&
\ln\frac{Q^2}{m_Q^2}
\left[ -\frac{1}{2}\int_0^1dz \frac{1+z^2}{1-z}
+ \int_y^1dz \frac{1}{1-z}
\right]
,\label{smodfin}
\\ 
S_q^{finite}(x,y)
&=&
-~\ln\left[(1-y)\frac{1-x}{x}\right]
- 2 - \ln(1-y)\ln(xy)
+~2 \, \litwo (1) - \litwo (xy)
\nl &&
+~\litwo
\left[-\frac{(1-x)y}{1-y}\right]
-
\left[
\litwo (1-y) + \litwo \left( - \frac{1-x}{x}\right)
\right]
{}.
\label{sfinite}
\ea

As is indicated by the choice of indices, the selected corrections may be
described by the following generic expression for
the collinear radiation of photons from the initial or final state
quark:
\ba
\frac{d^2 \sigma^{a}}{dxdy}
&=&
\frac{d^2 \sigma_{LLA,q}^{a}}{dxdy}
+
\sum_{Q,\bar Q}
\frac{2 \alpha^3 S x}{Q^4}
B_0(y,1)  f_Q(x)
c_q S_q^a(x,y|m_Q^2)
\\
\nonumber
&=&
\sum_{q=Q,\bar Q}
\frac{\alpha}{2\pi} Q_Q^2
\ln\frac{Q^2}{m_Q^2}
\int\limits_{0}^{1}dz
\frac{1+z^2}{1-z}
\left\{
\theta(z-z_0^a)
{\cal J}^a(x,y,Q^2)
\left.
\frac{d^2 \sigma_q^{0}}{dxdy}\right|
_{x={\hat x}, y={\hat y}, S={\hat
    S}}
-
\frac{d^2 \sigma_q^{0}}{dxdy}
\right\},
\label{lla1}
\ea
where $a=ini,fin$ and $m_Q$ is a constant quark mass.
The appropriate choices of initial and final state scalings are
\ba
{\hat p}_1 \stackrel{ini}{=} z p_1, \hspace{1.5cm}
{\hat p}_2 \stackrel{fin}{=} \frac{1}{z} p_2,
\label{scal}
\ea
and the corresponding Jacobian is
\ba
{\cal J}(x,y,Q^2) &=&
\left|\frac{\partial({\hat x},{\hat y})}{\partial(x,y)}\right|.
\label{Jcob}
\ea
Further details about scaling of kinematic variables
may be found in tables~\ref{LLAqi} and~\ref{LLAqf}.
These tables contain, in addition, the scaling properties for other
experimentally interesting sets of kinematical variables.

We now may address the following question:
Is it possible, as
has been proposed for the case of leptonic variables~\cite{Pet,Kripf,Spiesb},
to absorb the quark mass dependences completely by a redefinition of
the quark distribution functions?
For initial state radiation, this is evidently the case.
The LLA cross section in mixed variables has the same structure
as that in leptonic variables~\cite{disepl,Kripf}
and may be treated in the same way. Otherwise, the
contribution would be substantially overestimated (see ~\cite{Spiesb}).
This holds also for the other variables in table~1 with the notable
exception of the hadronic ones.

In leptonic variables, there is in accordance with the KLN theorem
no LLA contribution from final state radiation~\cite{KLN,disepl}.
For mixed variables (and other ones which are also not totally
inclusive in the hadronic final state (see table~2)), this is
different: There {\em is} a LLA contribution and its
Born function $B_0(y,z)$ in~(\ref{e4b})
does {\em not} factorize, preventing the framework as developed
in~\cite{Spiesb} to be operative here.

\vfill

\begin{table}[thbp]
{
\begin{center}
\begin{tabular}[]{|c|c|c|c|c|}
\hline
      \multicolumn{4}{|c}{}& \\
      \multicolumn{4}{|c}{\hspace{1.cm}Initial state radiation} & \\
      \multicolumn{4}{|c}{}& \\
\hline
          &&&&
\\
          &  $\hat{S}  $
          &  $\hat{Q}^2$
          &  $\hat{y}  $
          &  $z_0      $
\\
          &&&&              \\
\hline \hline
          &&&&                                              \\
leptonic  variables
    &  $ Sz                                             $
    &  $ \ql                                            $
    &  $ \yl                                            $
    &  $ \xl                                            $
\\
          &&&&                                              \\
\hline
          &&&&                                              \\
hadronic  variables
    &  $ Sz                                             $
    &  $\displaystyle{z \qh }                           $
    &  $\displaystyle{\yh   }                           $
    &  $ 0                                              $
\\
          &&&&
\\
\hline
 \end{tabular}
\end{center}
} 
\caption{\it
The definition of scaling variables for the
leading logarithmic corrections due to photon emission from the
initial quark leg.
Those variable sets of table~2, which are not contained
here transform as the leptonic variables do.
The definitions of variables are explained in detail in~[3,4].
\label{LLAqi}
}
\end{table}

\clearpage

\begin{table}[thbp]
{
\begin{center}
\begin{tabular}[]{|c|c|c|c|c|}
\hline
      \multicolumn{4}{|c}{}& \\
      \multicolumn{4}{|c}{\hspace{3cm} Final state radiation}& \\
      \multicolumn{4}{|c}{}& \\
\hline
          &&&&              \\
          &  $\hat{S}  $
          &  $\hat{Q}^2$
          &  $\hat{y}  $
          &  $z_0      $
\\
          &&&&              \\
\hline \hline
          &&&&                                              \\
leptonic  variables
    &  $ S                                              $
    &  $ \ql                                            $
    &  $ \yl                                            $
    &  $  0                                             $
\\
          &&&&                                              \\
\hline
          &&&&                                              \\
mixed     variables
    &  $  S                                             $
    &  $ \ql                                            $
    &  $ \displaystyle{\frac{\yh}{z} }                  $
    &  $ \yh                                            $
\\
          &&&&                                              \\
\hline
          &&&&                                              \\
hadronic  variables
    &  $ S                                              $
    &  $\displaystyle{\frac{\qh}{z}}                    $
    &  $\displaystyle{\frac{\yh}{z}}                    $
    &  $ \yh                                            $
\\
          &&&&                                              \\
\hline
          &&&&                                              \\
JB        variables
    &  $ S                                              $
    &  $ \displaystyle{\qjb \frac{1-\yh}{z(z-\yh)}}     $
    &  $ \displaystyle{\frac{\yh}{z}}                   $
    &  $ \displaystyle{\yh +\xjb(1-\yh)}                $
\\
          &&&&                                              \\
\hline
          &&&&                                              \\
double angle method
    &  $ S                                              $
    &  $ \qdo                                           $
    &  $ \ydo                                           $
    &  $  0                                             $
\\
          &&&&                                              \\
\hline
          &&&&                                              \\
$\theta_l, \yh$
    &  $ S                                              $
    &  $ \displaystyle{\qan \frac{z-\yh}
                      {z(1-\yh)}}                       $
    &  $ \displaystyle{\frac{\yh}{z}  }                 $
    &  $ \displaystyle{\frac{1-\yh(1-\xan)}{\xan}}      $
\\
          &&&&                                              \\
\hline
          &&&&                                              \\
$\Sigma $   method
    &  $ S                                              $
    &  $\displaystyle{\qsi\frac{[\ysi+z(1-\ysi)]}{z}}   $
    &  $\displaystyle{\frac{\ysi}{\ysi+z(1-\ysi)}}      $
   &  $ (*)                                                $
\\        &&&&
\\
\hline
          &&&&                                              \\
$e\Sigma $   method
    &  $ S                                              $
    &  $ \ql                                            $
    &  $\displaystyle{\frac{\yesi z}
                           {[\yesi+z(1-\yesi)]^2}}        $
   &  $ (*)                                                $
\\ &&&&
\\
\hline
 \end{tabular}
\end{center}
} 
\caption
{
{\it
The definition of scaling variables for the
leading logarithmic corrections due to photon emission from the final
state quark leg.
Variable sets without rescaling have no final state LLA radiation
(leptonic and double angle kinematics).
The definitions of variables are explained in detail in~[3,4].
}
\hfill
{\small
$^{(*)}$ $z_0 = [1 - 2xy (1-y) - (1 - 4xy (1-y))^{1/2}]
  / [2x (1-y)^2 ]$
}
\label{LLAqf}
}
\end{table}

\clearpage

\section{
\label{snum}
Numerical results}
%
The QED corrections will be described by the following correction factor:
\ba
\delta_{mix} &=& \frac{d^2 \sigma - d^2   \sigma^{Born}}
{d^2\sigma^{Born}} \times 100\%
\equiv \delta_e + \delta_i +\delta_q
{}.
\label{delta}
\ea
%
QED corrections in mixed variables are quite different from those
in leptonic variables.
Numerical results have been obtained with {\tt HECTOR}~\cite{Hector},
a Fortran program for the calculation of QED, electroweak, and QCD
corrections to
deep inelastic lepton nucleon scattering in different variables and
different formal approaches\footnote{
The parton distributions CTEQ2L~\cite{cteq} and
the {\tt HECTOR} flag {\tt ITERAD=1} are chosen.
This configuration uses the correct dependence of the structure
functions on $Q_h^2$.
We also include higher order LLA corrections
and the photonic vacuum polarization in the numerics. The latter gives
a sizeable contribution of 5 $\div$ 15 \%.
}:
{\tt LLA} -- leading logarithmic approximation, with possible inclusion of
higher order corrections;
{\tt MI} -- complete ${\cal O}(\alpha)$, plus (or without) higher order LLA,
[quark-parton] model-independent;
{\tt QPM} -- complete ${\cal O}(\alpha)$, plus (or without) higher order LLA,
in the quark-parton model.
The latter approach is advocated here.

As a matter of fact we mention that the leptonic corrections $\delta_e$
in mixed variables determined here in the {\tt QPM}
agree numerically exactly with those determined in the {\tt MI} (when using
the same structure functions).
The latter rely on completely different expressions and are described
in~\cite{MI}, where also the general
features of QED corrections in mixed variables are explained.

\begin{figure}[tbhp]
\begin{center}
\mbox
{
\epsfysize=9.cm
 \epsffile[160  40 470 490]{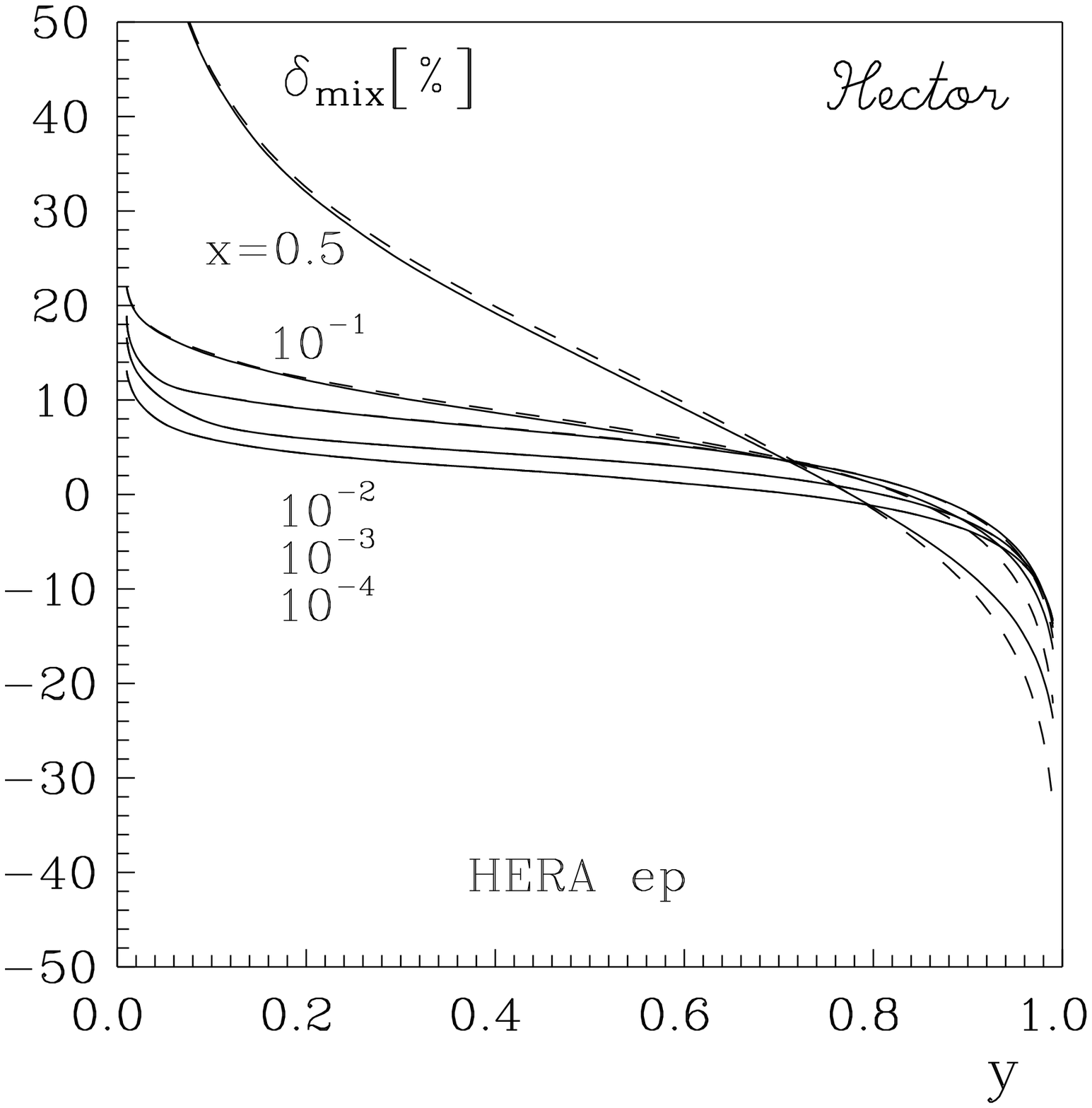}
}
\end{center}
\caption{\it
Comparison of the
radiative corrections $\delta_{mix}^{e+i}=\delta_e + \delta_i$
(broken line) with the leptonic corrections $\delta_e$ (solid line) at
HERA
in mixed variables.
\label{fdis6}
}
\end{figure}

\begin{figure}[tbhp]
\begin{center}
\mbox
{
\epsfysize=9.cm
 \epsffile[160  40 470 490]{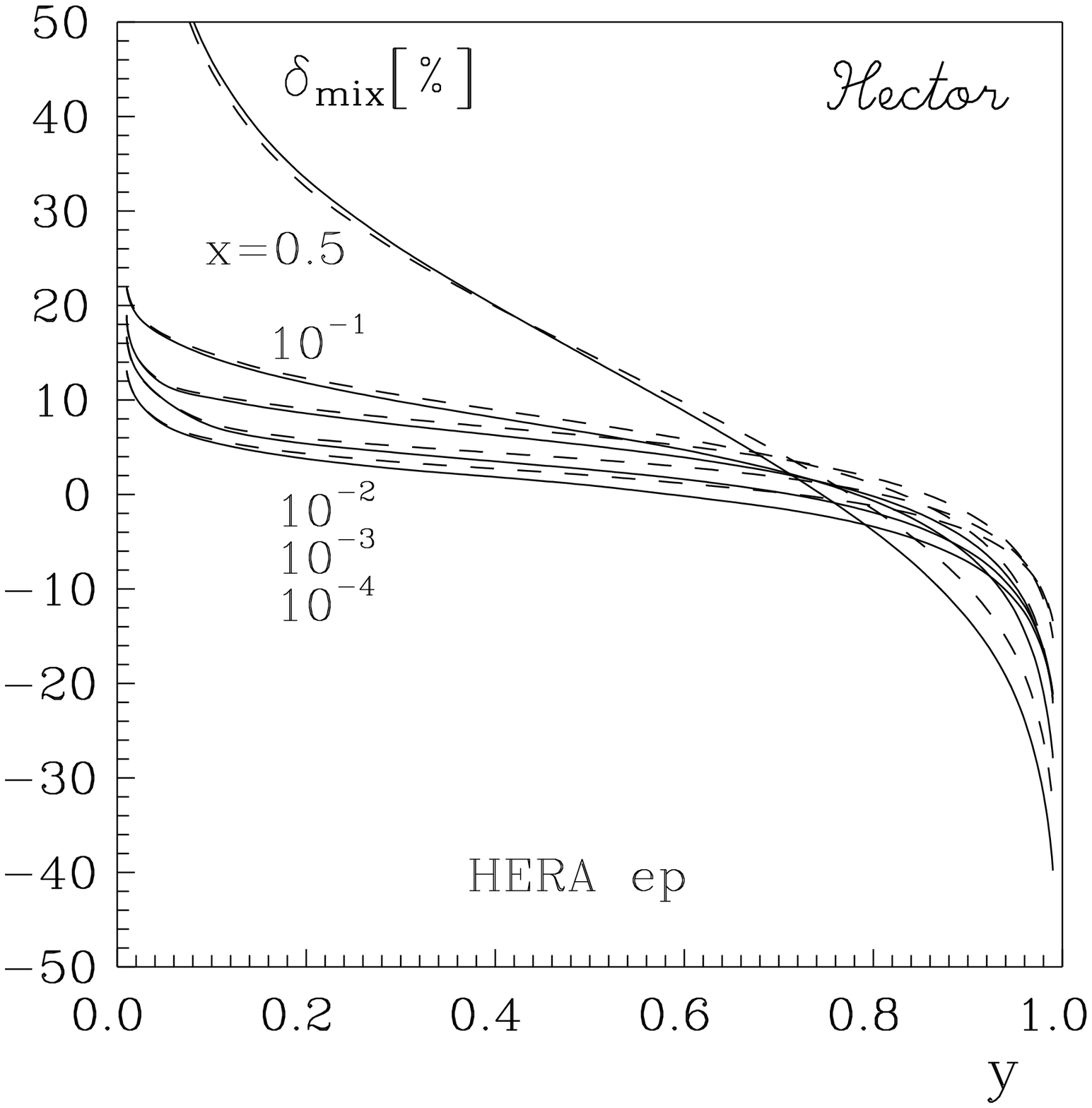}
}
\end{center}
\caption{\it
Comparison of $\delta_{mix}^{e+i}=\delta_e + \delta_i$
(broken line) with $\delta_{mix}=\delta_e + \delta_i +\delta_q$
(solid line) at HERA.
\label{fdis7}
}
\end{figure}

\begin{figure}[tbhp]
\begin{center}
\mbox
{
\epsfysize=9.cm
 \epsffile[160  40 470 490]{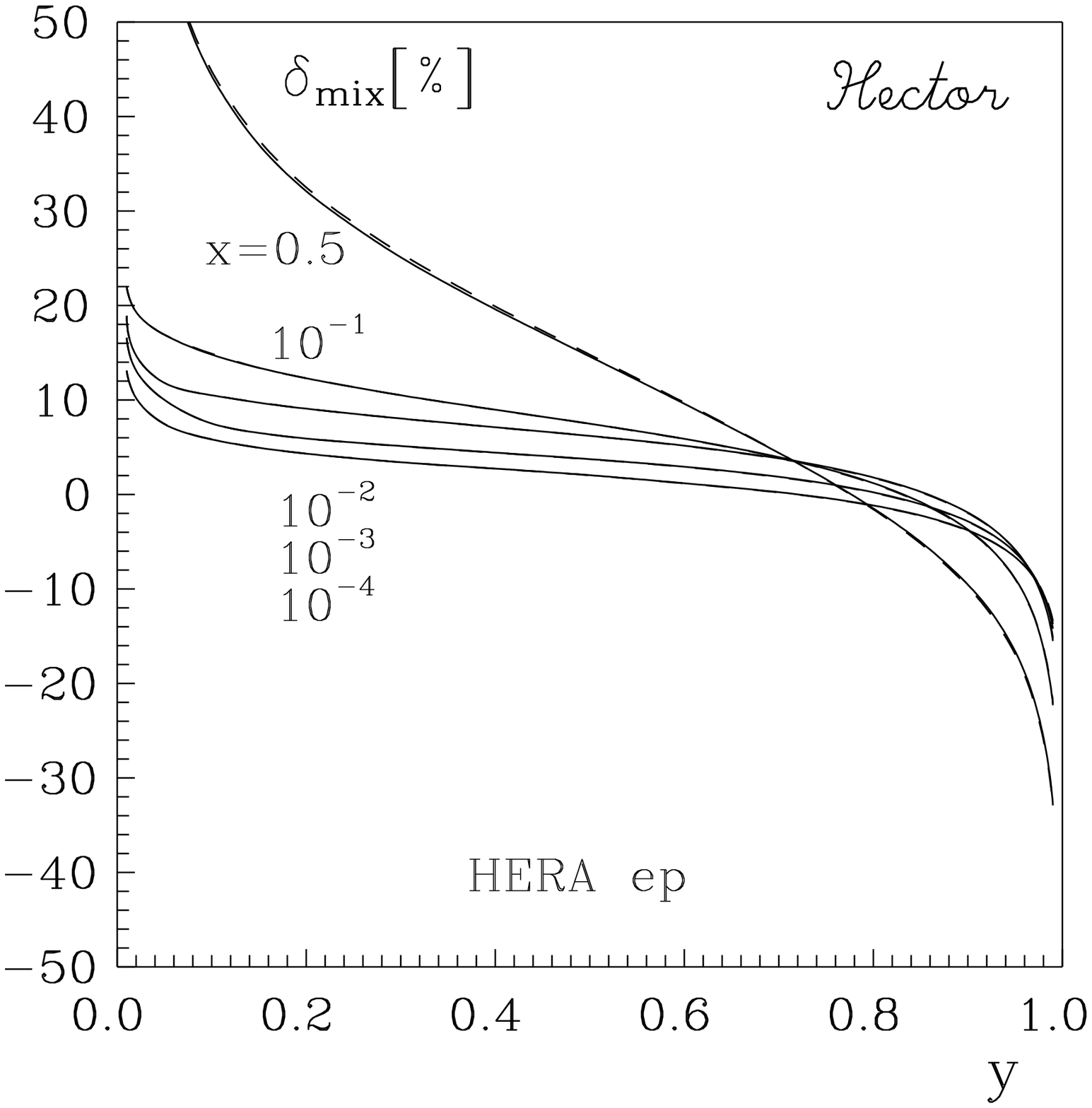}
}
\end{center}
\caption{\it
Comparison of $\delta_{mix}^{e+i}=\delta_e + \delta_i$
(broken line) with
$\delta_{mix}=\delta_e + \delta_i +\delta_q^{non-LLA}$
(solid line) at HERA.
\label{fdis8}
}
\end{figure}

On top of the leptonic corrections $\delta_e$,
which are the numerically largest
ones, the  effect of the interference corrections $\delta_i$
is shown in figure~\ref{fdis6}. 
They are positive at intermediate
values of $y$ and become negative at $y \geq 0.7$, nowhere exceeding
in magnitude 1\% except very large values of $x$ and $y$.
The additional influence of the quarkonic corrections is shown in
figures~\ref{fdis7} and~\ref{fdis8}.
In figure~\ref{fdis7} we present the effect of the quarkonic
corrections, calculated for definiteness with the assumption $m_Q = x_h M_p$.
Their effect is quite sizeable due to the unphysical quark mass
singularities.
Excluding both the initial and final state mass singularities from the
prediction (and assuming now the case of constant quark masses, which
chooses the sign discussed in footnote~4), we get figure~\ref{fdis8}.
In this case the quarkonic
corrections are rather small and the situation as shown in
figure~\ref{fdis6} is re-established.

The exclusion of initial state mass singularities is proposed
in~\cite{Spiesb}.
There it is shown that the photonic initial state LLA corrections,
although modifying the QCD evolution equations, give rise to
negligible numerical effects in the kinematical domain of HERA.
The exclusion of final state mass singularities might be motivated by
experimental reasons\footnote{We would like to thank J.~Bl\"umlein for
a clarifying discussion on this topic.}.
It is scarcely possible to distinguish between bremsstrahlung photons
and those from $\pi^0$ decay, say.
For an adequate treatment one has to redefine the kinematical
variables by adding the energy of a final state photon to the hadronic
energy.
In such variables, the final state LLA corrections would vanish in
accordance with KLN~theorem.

\bigskip
{\it To summarize,}
the QED corrections in mixed variables have a substantially different
behavior compared to those measured in leptonic variables.
The bulk of the corrections comes from the leptons.
If the data will reach an accuracy of the order of a percent one has
to take into account also the interference
bremsstrahlung corrections which, together with the quarkonic ones,
have been calculated here for the first time.
The major part of the latter is associated with quark mass singularities and
may be excluded from the corrections.

\bigskip
\noindent
{\bf Acknowledgment}
\\
We acknowledge the request of the referee to add section~3 on
quark mass singularities.
Two of us (P.~C. and L.~K.) would like to thank DESY -- Zeuthen for
the kind hospitality.


\end{document}